\documentclass[12pt]{iopart}
\usepackage{iopams}  
\usepackage{graphicx}
\usepackage{color}
\usepackage{subfigure}
\usepackage[dvipsnames]{xcolor}

\usepackage{tikz}
\usetikzlibrary{arrows,intersections}
\usepackage{pgfplots}
\usepackage[all]{xy}
\begin{document}

\title[Nests and Chains of Hofstadter Butterflies]{Nests and Chains of Hofstadter Butterflies}

\author{Indubala I. Satija$^{1}$ and Michael Wilkinson$^{2,3}$}

\address{
$^1$ Department of Physics, George Mason University, Fairfax, Virginia, USA,\\ 
$^2$ Chan Zuckerberg Biohub, 499 Illinois Street, San Francisco, CA 94158, USA,\\
$^3$ School of Mathematics and Statistics,
The Open University, Walton Hall, Milton Keynes, MK7 6AA, England
}
\vspace{10pt}
\ead{
isatija@gmu.edu, 
m.wilkinson@open.ac.uk
}
\begin{indented}
\item July 2019
\end{indented}

\begin{abstract}
The \lq Hofstadter butterfly', a plot of the spectrum of an electron 
in a two-dimensional periodic potential with a uniform magnetic 
field, contains subsets which resemble small, distorted images of the entire plot.
We show how the sizes of these sub-images are determined, and calculate scaling 
factors describing their self-similar nesting, revealing an un-expected simplicity in the 
fractal structure of the spectrum. We also characterise semi-infinite chains of sub-images, 
showing one end of the chain is a result of gap closure, and the other end is at an 
accumulation point.
\end{abstract}

\submitto{\jpa}
\maketitle

\section{Introduction}
\label{sec: 1}

\begin{figure}[h]
\centering
\includegraphics[width=0.9\textwidth]{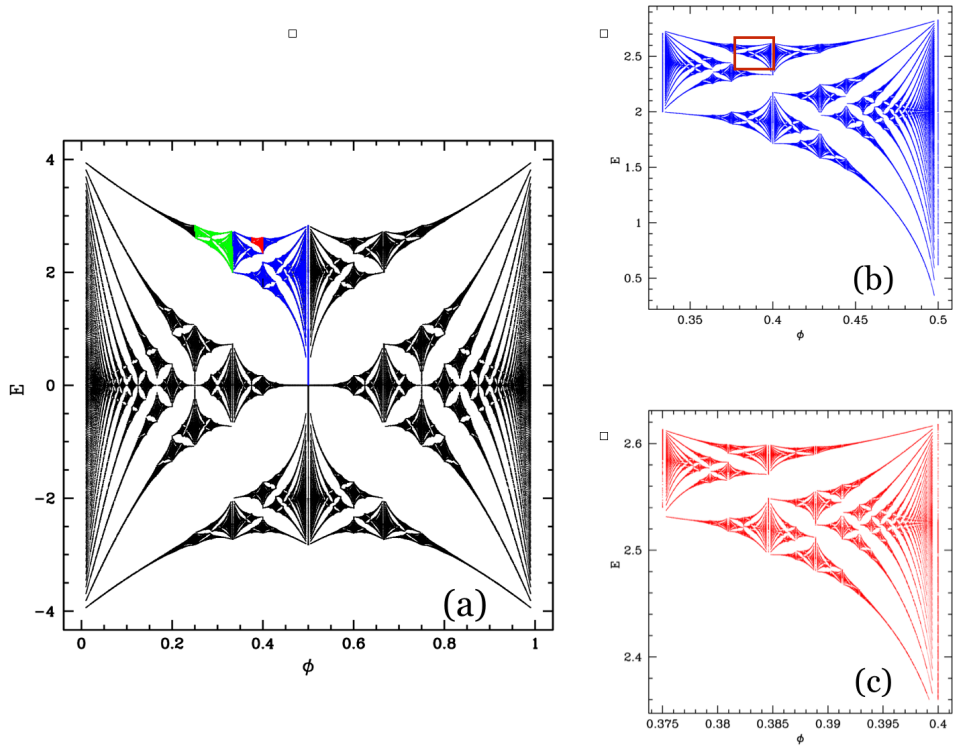}
\caption{
(Colour online). 
Illustrating Hofstadter's butterfly, and three sub-images. Because the red sub-image   
is nested inside the blue one, in that the red sub-image has the same relationship
to the blue one as the blue sub-image has to the whole plot, this nesting 
relationship can be made recursive, producing an infinite 
sequence of nested sub-images. The green and
blue sub-images form adjacent links in a chain. 
Panel (b) and (c) show the blowups of the blue and the 
red part of (a). After repeated application of this nesting, the approximate 
similarity  between the red and the blue approaches an exactly self-similar structure. 
}
\label{fig: 1}
\end{figure}

The \lq Hofstadter butterfly' \cite{Hof76} is a remarkable visual representation 
of the spectrum of Harper's equation \cite{Pei33,Har55}, a model for Bloch electrons
in a magnetic field. The butterfly plot, figure \ref{fig: 1}, shows allowed 
energy $E$ plotted vertically, as a function of a parameter $\phi$ 
which specifies the number of flux quanta per unit cell. The energies are 
plotted for values $\phi$ which are rational numbers $p/q$. 
The allowed energies consist of $q$ bands (with the central pair touching 
when $q$ is even). As remarked by Hofstadter \cite{Hof76}, 
the plot contains numerous smaller, distorted images of the 
whole pattern within it, largely confirming a prescient analysis of the 
problem made by Azbel' \cite{Azb64}. Three of these distorted sub--images  
are highlighted in figure \ref{fig: 1}. 
The edges of these sub-images occur at rational values of $\phi$, denoted 
by $\phi_{\rm R}$ and $\phi_{\rm L}$ (respectively, right and left edges), 
and  as described below, the locations of these edges follow simple 
number theoretical rules which are related \cite{Sat16} to the construction of the  
Farey tree \cite{Hardy}.

Figure \ref{fig: 1} also illustrates two ways in which the sub-images may be related to each 
other. One may be nested inside the other, such as the red sub-image nesting 
inside the blue one. These nesting relationships can be repeated recursively, and 
used to characterise the extent to which the pattern is self-similar. 
Another possibility is for sub-images to share a common vertical 
edge, such as the green sub-image sharing an edge with the blue one. The sub-images 
can be joined in this way to create chains. The successive sub-images in a chain become 
smaller as we follow the chain in one direction, and we shall see that the chains are 
semi-infinite, ending at an accumulation point. We find that the other end of the chain is 
a result of gaps in the spectrum closing.

This paper will describe the rules for determining which sub-sets of the butterfly
pattern can be classed as sub-images. We will show how these lead to a
variety of results about constructing nests and chains of sub-images, including 
the calculation of scaling factors quantifying the self-similarity of Hofstadter's plot.

The tool for this investigation will be a renormalisation-group analysis 
of the spectrum,  originally described in \cite{Wil87} (a more elegant 
formulation was subsequently made possible using generalised Wannier functions \cite{Wil98,Wil00}). 
Section \ref{sec: 2} describes the principle result on renormalisation of $\phi$ from 
\cite{Wil87} which is key to obtaining our results, together with new formulae for 
renormalisation of the quantised Hall conductance integers. 

Section \ref{sec: 3} will discuss how to label a sub-image by specifying one of 
its edges, and how to determine the flux value $\phi$ for its other edge and for its 
centre. Recently, based upon numerical investigations, it has been proposed  
that these sub-images may be generated from a partition of the plot which is closely 
related to the Farey tree construction, and to other number-theoretical constructions, 
including Ford circles, integral Apollonian packings \cite{Sat16,SatEP16} and 
Pythagorean triplets \cite{Sat18}. The results in section \ref{sec: 3} explain the connection 
of the sub-images to Farey trees.

Section \ref{sec: 4} will consider how sub-images can be nested in a 
self-similar way, leading to exact expressions for scaling factors describing 
self-similar nesting. Some of these results were previously obtained by 
empirical observations, guided by observations of connections to Farey trees 
\cite{Sat16,SatEP16}. Section \ref{sec: 5} discusses the construction of chains of sub-images. 
We show that they are infinite in one direction, ending in an accumulation 
point at a rational value of $\phi$, and discuss how the chains terminate 
at the other end due to closure of gaps in the spectrum. Our 
conclusions are confirmed by numerical illustrations
throughout. Finally section \ref{sec: 6} presents a summary our findings. 

The Hofstadter butterfly plot has stimulated investigations which 
use a vast range of different methods, for example \cite{Wie+94} and \cite{Las94} 
are significant contributions which use very different approaches from our work. There 
is an extensive survey of the literature in \cite{Sat16}. However, we are aware 
of one other study which emphasises making a partition of the Hofstadter 
butterfly plot: Osadchy and Avron \cite{Osa+01} treat the Hofstadter plot as 
a phase-diagram, where the phases are labelled by quantum Hall conductances. 
Their approach is complementary to our own, in that their partition is based 
upon labelling gaps in the spectrum, rather than partitioning the spectrum itself.

\section{Results of renormalisation group analysis}
\label{sec: 2}

Although many aspects of the Hofstadter butterfly plot \cite{Hof76} are singularly 
discontinuous, the gaps in the spectrum of the Harper equation Hamiltonian 
$\hat H(\phi)$ are stable features, which are continuous 
under variation of the flux parameter $\phi$. The renormalisation-group method 
exploits this observation, by relating the complex spectrum at a typical value 
of $\phi$ to the much simpler spectrum at a nearby rational value, $\phi=p_0/q_0$.

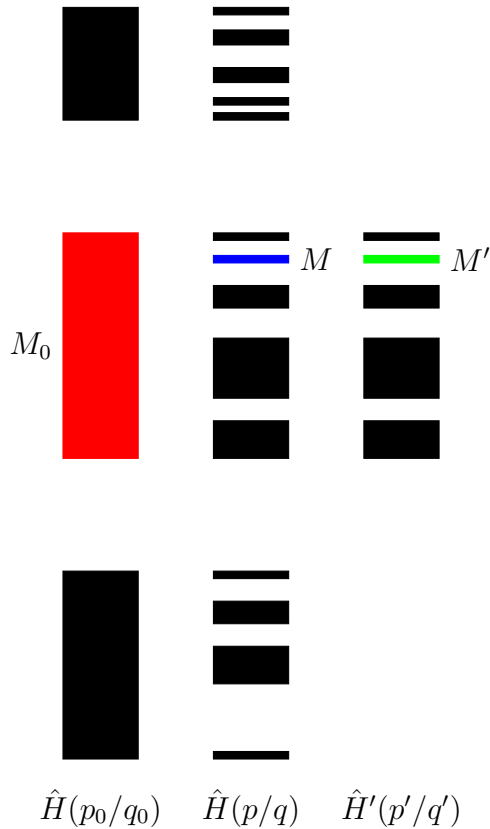
\begin{figure}[t!]
\centering
\bigskip 
\begin{tikzpicture}

\draw [fill=black,black] (0.0,0.0) rectangle (1.0,2.5);
\draw [fill=red,red] (0.0,4.0) rectangle (1.0,7.0);
\draw [fill=black,black] (0.0,8.5) rectangle (1.0,10.0);

\draw [fill=black,black] (2.0,0.0) rectangle (3.0,0.1);
\draw [fill=black,black] (2.0,1.0) rectangle (3.0,1.5);
\draw [fill=black,black] (2.0,1.8) rectangle (3.0,2.1);
\draw [fill=black,black] (2.0,2.4) rectangle (3.0,2.5);

\draw [fill=black,black] (2.0,4.0) rectangle (3.0,4.5);
\draw [fill=black,black] (2.0,4.8) rectangle (3.0,5.6);
\draw [fill=black,black] (2.0,6.0) rectangle (3.0,6.3);
\draw [fill=blue,blue] (2.0,6.6) rectangle (3.0,6.7);
\draw [fill=black,black] (2.0,6.9) rectangle (3.0,7.0);

\draw [fill=black,black] (2.0,8.5) rectangle (3.0,8.6);
\draw [fill=black,black] (2.0,8.7) rectangle (3.0,8.8);
\draw [fill=black,black] (2.0,9.0) rectangle (3.0,9.2);
\draw [fill=black,black] (2.0,9.5) rectangle (3.0,9.7);
\draw [fill=black,black] (2.0,9.9) rectangle (3.0,10.0);

\draw [fill=black,black] (4.0,4.0) rectangle (5.0,4.5);
\draw [fill=black,black] (4.0,4.8) rectangle (5.0,5.6);
\draw [fill=black,black] (4.0,6.0) rectangle (5.0,6.3);
\draw [fill=green,green] (4.0,6.6) rectangle (5.0,6.7);
\draw [fill=black,black] (4.0,6.9) rectangle (5.0,7.0);

\node [left] at (0.0,5.5) {$M_0$};
\node [right] at (3.0,6.65) {$M$};
\node [right] at (5.0,6.65) {$M'$};
\node [below] at (0.5,-0.25) {$\hat H(p_0/q_0)$};
\node [below] at (2.5,-0.25) {$\hat H(p/q)$};
\node [below] at (4.5,-0.25) {$\hat H'(p'/q')$};

 \end{tikzpicture}
\caption{
(Colour online). 
When $\phi$ is close to $\phi_0=p_0/q_0$, the spectrum of the Hamiltonian $\hat H(\phi)$ 
divides into $q_0$ bands. The renormalisation-group method constructs an effective 
Hamiltonian $\hat H'$, such that $\hat H'(\phi')$ has a spectrum which is 
equal to the subset of the spectrum of $\hat H$ contained in one of these bands.

In this case we construct $\hat H'(\phi')$ for a band (highlighted in red) 
having Hall conductance integer $M_0$. We are interested in the case 
where $\phi=p/q$ (and consequently $\phi'=p'/q'$) 
are rational numbers, so the spectra are sets of bands. The band highlighted 
in blue has Hall conductance integer $M$, and it corresponds to a band of $\hat H'(\phi')$ 
(highligted in green) with Hall conductance $M'$.
} 
\label{fig: 2}
\end{figure}

Inspection of the plot shows that, when $\phi$ is close to the rational 
value $\phi_0=p_0/q_0$, the spectrum clusters into $q_0$ regions, which correspond 
with the the $q_0$ bands of the spectrum when $\phi=p_0/q_0$. 
In \cite{Wil87}, it is shown that 
for small values of $\phi-\phi_0$, each of the $q_0$
bands is transformed into the spectrum of a renormalised 
effective Hamiltonian operator $\hat H'(\phi')$, which 
has a structure which is analogous to the original 
problem, with a renormalised value of $\phi$, denoted by $\phi'$. 
The transformation is a renormalisation 
group because it eliminates degrees of freedom: the transformed effective 
Hamiltonian only describes the spectrum of one of $q_0$ bands in the spectrum 
of the original Hamiltonian. 
In this paper we shall usually be concerned with cases 
where $\phi$ is also a rational, equal to $p/q$. In this case the renormalised 
Hamiltonian also has a rational value of the flux parameter: $\phi'=p'/q'$, 
where $p'$ and $q'$ are integers. 
The relationship between between the original 
Hamiltonian $\hat H(\phi)$ and the renormalised Hamiltonian $\hat H'(\phi')$ is summarised 
in figure \ref{fig: 2}.

The Harper equation can be viewed as a model for either the perturbation of a Landau level 
by a periodic potential \cite{Sch+81,Wil87a}, or as a model for a Bloch band 
perturbed by a magnetic field \cite{Pei33,Rot62}. In the former
case $\phi$ is the ratio of the area of the flux quantum to the area $A$ of the unit cell, that is 
$\phi=h/eBA$, and in the latter case $\phi$ is the reciprocal of this quantity. We shall adopt 
the perturbed Landau level picture for the discussion in this paper. The total density of states 
in the Landau level is $eB/h$. When $\phi$ is rational, the $q$ different bands of the spectrum 
each has a quantised Hall conductance
\begin{equation}
\label{eq: 2.1}
\sigma^{(k)}=\frac{e^2}{h} M^{(k)}
\end{equation}
where $M^{(k)}$ is the quantised Hall conductance integer of the band with index $k$ \cite{Tho+81}.
The total Hall conductance of a single Landau level is $\sigma^{({\rm LL})}=e^2/h$, so that
\begin{equation} 
\label{eq: 2.2}
\sum_{k=1}^q M^{(k)}=1
\ .
\end{equation}

\subsection{Renormalisation of $\phi$}
\label{sec: 2.1}

In \cite{Wil87} it is shown that the renormalised effective Hamiltonian $\hat H'$
has a renormalised value of the flux parameter $\phi$ which is given by
\begin{equation}
\label{eq: 2.1.1}
\phi'=\frac{q_0\phi-p_0}{N_0\phi+M_0}
\end{equation}
where $M_0$ is the quantised Hall conductance integer of the band
and $N_0$ is another integer, satisfying
\begin{equation}
\label{eq: 2.1.2}
1=q_0M_0+p_0N_0
\ .
\end{equation}

The structure of the renormalised Hamiltonian is specified by a set 
of Fourier coefficients, $H'_{nm}$ \cite{Wil87,Wil+96}. The evaluation of these coefficients 
is complicated, but analytical approximations are available when $\phi$ 
is small \cite{Sus82,Wil84}, and when $\phi'$ is small \cite{Wil+96}. For the 
purposes of this paper, however, we only use the fact that the renormalised 
Hamiltonian has a spectrum which closely resembles that of the original 
Hamiltonian, for a different value of $\phi$, and with the energies subjected 
to a linear transformation, as discussed in \cite{Sus82,Wil84,Wil+96}.

\subsection{Renormalisation of $M$}
\label{sec: 2.2}

Because the expression for renormalisation of $\phi$ (equation (\ref{eq: 2.1.1})) 
depends upon the values of the Hall conductance integers $M_0$ and $N_0$, 
if we are to iterate the transformation our analysis requires information about
how these integers change as the renormalisation scheme is iterated.
In general the Hall conductance integers $M$ can be determined by numerically 
computing the Chern index of the band, as described by 
Thouless  {\sl et al.} \cite{Tho+81}, but this is 
cumbersome and un-informative. Instead, the recursive structure of the 
spectrum can be associated with a recursive method to 
compute the Hall conductance $M$ of a given band.
The argument was not presented in the earlier works 
\cite{Wil87,Wil98,Wil00} describing the renormalisation scheme. 

Specifically, we wish to determine the Hall conductance integer 
$M$ of a given band when $\phi=p/q$, such as the one highlighted 
in blue in figure \ref{fig: 2}. We shall assume that 
this band lies within a cluster of bands corresponding to a band of 
the spectrum when $\phi=p_0/q_0$, having Hall conductance integer $M_0$. 
This cluster of bands is described by an effective Hamiltonian $\hat H'$ with $q'$ bands, 
and the band highlighted in green corresponds to a band of this Hamiltonian, with 
Hall integer $M'$. In order to probe the structure of the spectrum by 
recursive application of the renormalisation group transformation, it would 
be useful to be able to express the integer $M$ in terms 
of $M_0$, $M'$ and possibly other integers.

This can be achieved using the Str\v{e}da formula \cite{Str81}, according to which the Hall 
conductance when the Fermi level is in a gap is 
\begin{equation}
\label{eq: 2.2.1}
\sigma=e\frac{\partial {\cal N}}{\partial B}
\end{equation}
where ${\cal N}$ is the number of filled states per unit area below the gap. 
We chose to interpret the Harper equation as representing the 
perturbation of a Landau level by a periodic potential, which
has electron density $eB/h$. Then the quantity $\phi$ is the ratio of the area of a
flux quantum to the area $A$ of the unit cell: $\phi=h/eBA$. 
If the filling fraction of the states below the band gap 
is $\nu$, we then find that 
\begin{equation}
\label{eq: 2.2.2}
\sigma=\frac{e^2}{h}\left[ \nu-\phi \frac{\partial \nu}{\partial \phi}\right]
\ .
\end{equation}
The \lq gap-labelling theorem' (Claro and Wannier \cite{Cla+79}) 
implies that 
\begin{equation}
\label{eq: 2.2.3}
\nu=\bar N\phi+\bar M
\end{equation}
for some integers $\bar N$ and $\bar M$. Substitution of (\ref{eq: 2.2.3}) into 
(\ref{eq: 2.2.2}) shows that $\sigma=\bar Me^2/h$, so that $\bar M$ is the Hall
conductance integer. If the spectrum is separated into several bands, we can 
write the filling fraction for each band in the same form as (\ref{eq: 2.2.3}):
\begin{equation}
\label{eq: 2.2.3a}
\Delta \nu=N\phi +M
\end{equation}
where $M$ is the quantised Hall conductance integer associated with a band, 
and $N$ is a complementary integer.

We assume that it is possible to determine 
the Hall conductance integers when the denominator $q$ of the flux ratio $\phi=p/q$ is small. Let us 
approximate $\phi$ by $\phi_0=p_0/q_0$, where $q_0$ is sufficiently 
small that we can readily determine the Hall integers $M_0$, 
for all of its bands, together with the conjugate 
integers $N_0$ satisfying $p_0N_0+q_0M_0=1$. We implement 
the renormalisation procedure of \cite{Wil87} using the rational 
$p_0/q_0$ as the base case. We shall determine the Hall conductance integer
from the spectrum alone by using the Str\v{e}da formula in the form (\ref{eq: 2.2.2}).

The band for which we wish to determine the Hall conductance integer is 
in a cluster of bands for the $\phi_0$ commensurability. The filling fraction 
of this cluster is denoted by $\Delta \nu_0$. The band that we are interested 
in lies within this cluster, and its filling fraction relative to all of the states 
of the cluster is $\Delta \nu'$. The overall filling fraction of the sub-band, 
which will be used to determine the Hall conductance integer, is 
\begin{equation}
\label{eq: 2.2.4}
\Delta \nu(\phi)=\Delta \nu_0\ \Delta \nu'
\ .
\end{equation}
The filling fractions are obtained by applying (\ref{eq: 2.2.3a}) 
to the original problem and to the renormalised problem in turn. 
The renormalised Hamiltonian has commensurability $\phi'=p'/q'$, 
and the band that we wish to analyse corresponds to a sub-band of the 
renormalised Hamiltonian, with gap-labelling integers $N'$ and $M'$.
The filling factors in (\ref{eq: 2.2.4}) are, therefore, 
\begin{eqnarray}
\label{eq: 2.2.5}
\Delta \nu_0&=&N_0\phi+M_0
\nonumber \\
\Delta \nu'&=&N' \phi'+M'
\ .
\end{eqnarray}
The renormalised commensurability is given by equation (\ref{eq: 2.1.1}). 
Substituting (\ref{eq: 2.2.5}) into (\ref{eq: 2.2.4}) and using (\ref{eq: 2.1.1}) yields
\begin{equation}
\label{eq: 2.2.7}
\Delta \nu =(q_0N'+N_0M')\phi +(M_0M'-p_0N')
\ .
\end{equation}
Comparison with (\ref{eq: 2.2.3a}) shows that the equations for renormalisation of the Hall
conductance integers are
\begin{equation}
\label{eq: 2.2.8}
M=M_0M'-p_0N'
\ ,\ \ \ 
N=N_0M'+q_0N'
\ .
\end{equation}

\section{Labelling and construction of sub-images: relation to Farey trees}
\label{sec: 3}

Note that, when $\phi=0$ or when $\phi=1$, the spectrum 
of the Harper equation is a single interval (which happens to be $[-4,4]$). 
This indicates that we can define \lq sub-images' by identifying one edge with 
a single band of the spectrum, bounded by a gap on either side. For example, 
we may take one particular band of the spectrum when $\phi_0=p_0/q_0$ as 
forming the left-hand edge of the sub-image. As we increase $\phi$ away from 
its initial value of $\phi_{\rm L}=p_{\rm L}/q_{\rm R}$, 
the spectrum becomes very complex, but the gaps 
which separate the sub-image from the rest of the spectrum persist. We may 
find that when $\phi$ reaches another rational value, $\phi_{\rm R}=p_{\rm R}/q_{\rm R}$, 
the complex spectrum re-forms into a single band, with the same open gaps above and below. 
Furthermore, the structure of the sub-spectrum 
in the region between $\phi_{\rm L}$ and $\phi_{\rm R}$ is well-approximated by a 
distorted version of the original Hofstadter butterfly. 

In order to describe the sub-images we first need a convention for labelling them. 
We label the sub-image by specifying an energy band that forms either 
the left-hand or the right-hand edge. Thus a sub-image is specified 
using a rational number, either $\phi_{\rm L}=p_{\rm L}/q_{\rm L}$ or 
$\phi_{\rm R}=p_{\rm R}/q_{\rm R}$, depending on 
whether we construct the sub-image by starting from its left or right edge, respectively. 
We must also specify the index of the band (labelling them consecutively in increasing 
energy). The band index will be denoted by $k$. According to this convention, the 
sub-images highlighted in figure \ref{fig: 1} are:
\begin{eqnarray}
\label{eq: 3.1}
{\rm blue}&:&\phi_{\rm L}=1/3\ ,\ \ \ k=3\ ,\ \ \ M=0\ ,\ \ \ N=1
\nonumber \\
{\rm red}&:&\phi_{\rm L}=3/8\ ,\ \ \ k=8\ ,\ \ \ M=-1\ ,\ \ \ N=3
\nonumber \\
{\rm green}&:&\phi_{\rm L}=1/4\ ,\ \ \ k=4\ ,\ \ \ M=0\ ,\ \ \ N=1
\ .
\end{eqnarray}
In these cases the corresponding values of the opposite edges 
of the sub-images are, respectively,
\begin{equation}
\label{eq: 3.2}
\phi_{\rm R}=1/2\ ,\ \ \ \phi_{\rm R}=2/5\ ,\ \ \ \phi_{\rm R}=1/3  
\ .
\end{equation}
Given the value of $\phi_{\rm L}$, we determine $\phi_{\rm R}$ 
by the requirement that the \emph{renormalised} value of $\phi$
is equal to unity at the opposite edge. Inverting  (\ref{eq: 2.1.1})
we obtain
\begin{equation}
\label{eq: 3.3}
\phi=\frac{p_0+M_0\phi'}{q_0-N_0\phi'}
\ .
\end{equation}
If we start at $\phi_{\rm L}$ (the left-hand edge), the 
other edge of the sub-image is obtained by 
setting $\phi'=1$, so that on setting $\phi_0=\phi_{\rm L}$, equation (\ref{eq: 3.3}) 
implies that $\phi_{\rm R}$ is given by
\begin{equation}
\label{eq: 3.4}
\phi_{\rm R}\equiv\frac{p_{\rm R}}{q_{\rm R}}=
\frac{p_{\rm L}+M_0}{q_{\rm L}-N_0}
\ .
\end{equation}
The centre of the sub-image between $\phi_{\rm L}$ and $\phi_{\rm R}$ is at a rational 
value $\phi_{\rm c}=p_{\rm c}/q_{\rm c}$.
The centre of the sub-image is defined by setting $\phi'=\frac{1}{2}$, 
so that $\phi_{\rm c}$ is given by
\begin{equation}
\label{eq: 3.5}
\phi_{\rm c}\equiv\frac{p_{\rm c}}{q_{\rm c}}
=\frac{M_0+2p_{\rm L}}{2q_{\rm L}-N_0}
\ .
\end{equation}
Thus equations (\ref{eq: 3.4}) and (\ref{eq: 3.5}) show that 
$\phi_{\rm c}$ is not the arithmetic mean of $\phi_{\rm L}$ and $\phi_{\rm R}$, but 
rather 
\begin{equation}
\label{eq: 3.6}
\frac{p_{\rm c}}{q_{\rm c}}=\frac{p_{\rm L}+p_{\rm R}}{q_{\rm L}+q_{\rm R}}
\ .
\end{equation}
This is the \lq Farey sum' of the two edge values. 
Furthermore  the equations (\ref{eq: 3.4}) and (\ref{eq: 3.6}) imply that
\begin{equation}
\label{eq: 3.61}
|q_{\rm L} p_{\rm R} - q_{\rm R} p_{\rm L} | = 1,\,\,\ |q_{\rm L} p_{\rm c} - q_{\rm c} p_{\rm L} | 
= 1,\,\,\ |q_{\rm R} p_{\rm c} - q_{\rm c} p_{\rm R} | = 1
\ .
\end{equation}
This means that the triplets $[\frac{p_{\rm L}}{q_{\rm L}}, \frac{p_{\rm c}}{q_{\rm c}}, \frac{p_{\rm R}}{q_{\rm R}}]$  
which are the flux values at the edges and at the 
centres of the sub-images are \lq neighbours'  in the 
Farey tree. Such triplets of rational numbers are also known as
\emph{friendly numbers}.  Empirical evidence supporting (\ref{eq: 3.6}) and 
(\ref{eq: 3.61}) was previously discussed in  \cite{Sat16,SatEP16}. Figure \ref{fig: 3} below 
illustrates the application of equations (\ref{eq: 3.4}) and (\ref{eq: 3.6}).

\section{Nesting of sub-images} 
\label{sec: 4}

\subsection{Recursive nesting}
\label{sec: 4.1}

The sub-images in the Hofstadter butterfly plot can be nested recursively. 
Here we consider how to describe and quantify this. The nested 
sub-images converge to a fixed point, and we can determine exact expressions 
for the scaling factors describing the ratio of size of different levels for the 
hierarchy. The fact that exact expressions for the scaling ratios can be determined 
is a little surprising. The crucial ingredient is that the scaling factors do not depend 
upon the structure of the Hamiltonian, only upon the explicit expressions for renormalisation 
of $\phi$ and of the Hall integers, $M$ and $N$.

We consider the following recursion. Pick a left band edge specifying a sub-image. 
This is described by integers $p^{\rm L}_0$, $q^{\rm L}_0$, $M_0$ and $N_0$. We then consider 
a next-generation sub-image, which is obtained by picking a sub-image 
with the \lq internal' coordinate of the left band edge at $\phi'=\tilde p^{\rm L}/\tilde q^{\rm L}$, 
with Hall conductance integers of the renormalised Hamiltonian equal to 
$\tilde M$ and $\tilde N$. For example, in figure \ref{fig: 1}, the blue sub-image 
corresponds to $p^{\rm L}_0=1$, $q^{\rm L}_0=3$, $M_0=0$, $N_0=1$, and the red sub-image 
nested inside corresponds to the same values (that is $\tilde p^{\rm L}=1$, $\tilde q^{\rm L}=3$, 
$\tilde M=0$, $\tilde N=1$).

\begin{figure}[h]
\centering
\includegraphics[width=0.95\textwidth]{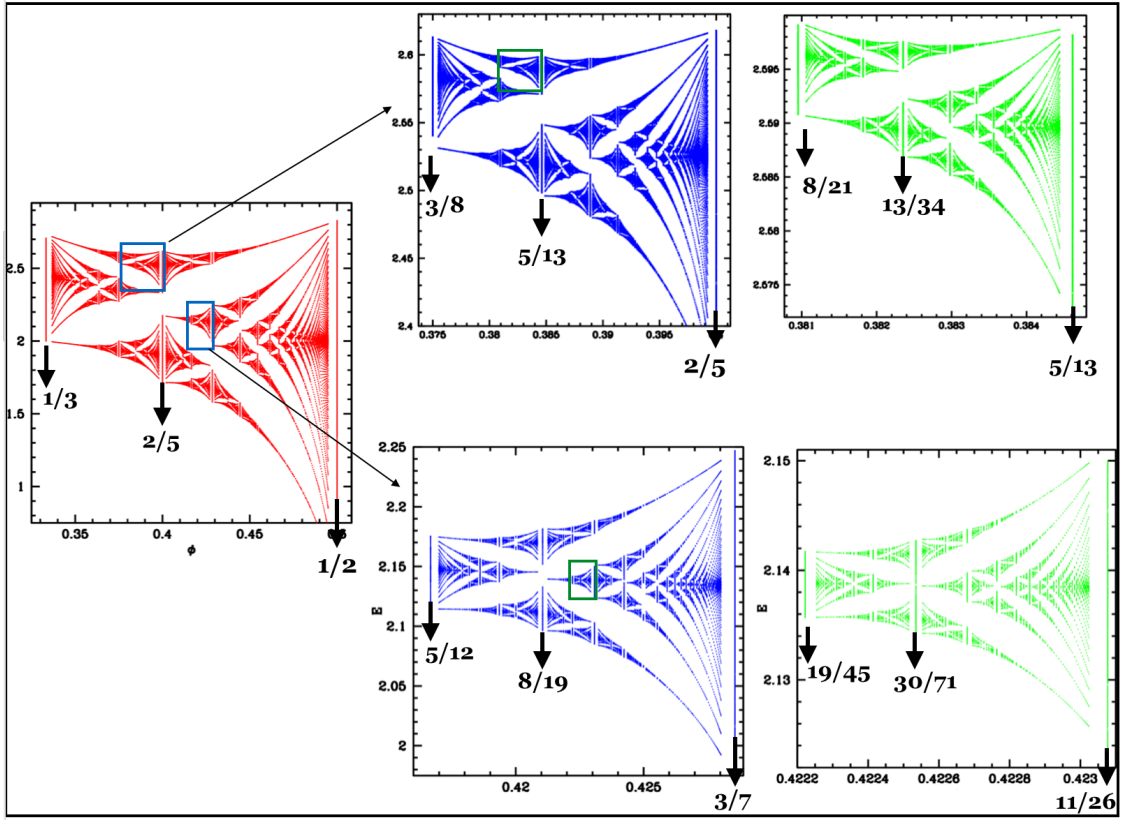}
\caption{
(Colour online).
Illustrating  three generations of two distinct nested  
sequences of sub-images that appear in the blue sub-image from figure \ref{fig: 1}.  
The values of $\phi_j=p_j/q_j$ at the left-hand edge are predicted 
using equations (\ref{eq: 4.3}) and (\ref{eq: 4.4}): the values of the coefficients 
for this example are discussed in section \ref{sec: 4.3}. The values of $\phi$ 
at the right-hand edge and at the centre of each sub-image are then obtained 
using equations (\ref{eq: 3.4}) and (\ref{eq: 3.6}) respectively. 
}
\label{fig: 3}
\end{figure}

We can then repeat this construction recursively, so that at stage $j$, the left band edge
is $\phi^{\rm L}_j=p^{\rm L}_j/q^{\rm L}_j$, and the Hall integers are $M_j$ and $N_j$. At the next stage 
of the iteration, we determine a new band edge with its left hand edge at a \emph{fixed} value of the 
\emph{renormalised} flux parameter, $\phi'$, equal to {$\tilde p^{\rm L}/\tilde q^{\rm L}$. The iteration of 
the $\phi^{\rm L}_j$ is therefore given by applying (\ref{eq: 3.3}), 
replacing the $\phi^{\rm L}_0=p^{\rm L}_0/q^{\rm L}_0$ 
with the flux ratio of the left-hand band edge at level $j$ of the nesting:
\begin{equation}
\label{eq: 4.1}
\phi^{\rm L}_{j+1}=\frac{p^{\rm L}_{j+1}}{q^{\rm L}_{j+1}}
=\frac{p^{\rm L}_j\tilde q^{\rm L}+M_j\tilde p^{\rm L}}{q^{\rm L}_j\tilde q^{\rm L}-N_j\tilde p^{\rm L}}
\ .
\end{equation}
The iteration of the Hall integers is obtained from (\ref{eq: 2.2.8}): changing to 
the notation of the current application, replacing $(M,N)$ with $(M_j,N_j)$, these read
\begin{equation}
\label{eq: 4.2}
M_{j+1}=M_j\tilde M-p^{\rm L}_j\tilde N
\ ,\ \ \ 
N_{j+1}=N_j\tilde M+q^{\rm L}_j\tilde N
\ .
\end{equation}

These recursion equations describing the left-hand edges of the sub-images 
are a linear system of the form 
$\mbox{\boldmath$x$}^{\rm L}_{j+1}={\bf A}^{\rm L}\,\mbox{\boldmath$x$}^{\rm L}_j$:
\begin{equation}
\label{eq: 4.3}
\left(\begin{array}{c}
p^{\rm L}_{j+1} \cr
M_{j+1} \cr
q^{\rm L}_{j+1} \cr
N_{j+1}
\end{array}\right)
=
{\bf A}^{\rm L}
\left(\begin{array}{c}
p^{\rm L}_j \cr
M_j \cr
q^{\rm L}_j \cr
N_j
\end{array}\right)
\end{equation}
where
\begin{equation}
\label{eq: 4.4}
{\bf A}^{\rm L}=
\left(\begin{array}{cccc}
\tilde q^{\rm L}&\tilde p^{\rm L}&0&0 \cr
-\tilde N&\tilde M&0&0 \cr
0&0&\tilde q^{\rm L}&-\tilde p^{\rm L} \cr
0&0&\tilde N&\tilde M
\end{array}\right)
\ .
\end{equation}
This representation is convenient because the left hand edges of the 
nested bands are represented by a vector $\mbox{\boldmath$x$}_j^{\rm L}=(p_j^{\rm L},M_j,q^{\rm L}_j,N_j)$}.
We observe that equations (\ref{eq: 4.3}) and (\ref{eq: 4.4}) can be expressed in a more elegant
and symmetrical form using $2\times 2$ matrices: 
\begin{equation}
\label{eq: 4.4a}
{\bf B}^{\rm L}_j=
\left(\begin{array}{cc}
q_j^{\rm L} & p_j^{\rm L} \cr
-N_j & M_j
\end{array}\right)
\ ,\ \ \ 
{\bf C}^{\rm L}=
\left(\begin{array}{cc}
\tilde q^{\rm L} & \tilde p^{\rm L} \cr
-\tilde N & \tilde M
\end{array}\right)
\ .
\end{equation}
Note that, as a consequence of (\ref{eq: 2.1.2}), these matrices are unimodular: 
${\rm det}({\bf B}_j^{\rm L})={\rm det}({\bf C}^{\rm L})=1$.
With these definitions, equations (\ref{eq: 4.3}) and (\ref{eq: 4.4})  are equivalent 
to a matrix multiplication process:
\begin{equation}
\label{eq: 4.4b}
{\bf B}^{\rm L}_{j+1}={\bf C}^{\rm L} {\bf B}^{\rm L}_j 
\ .
\end{equation}
Because any unimodular $2\times 2$ matrix ${\bf C}$ satisfies 
${\bf C}^2+{\bf I}={\rm tr}({\bf C}){\bf C}$ (where ${\bf I}$ is the identity matrix), 
equation (\ref{eq: 4.4b}) can be rewritten as a set of four two-term recursions in which 
the variables $(p_j,M_j,q_j,N_j)$ are decoupled:
\begin{equation}
\label{eq: 4.4c}
s_{j+2} = ( \tilde q^L + \tilde M) s_{j+1} - s_j
\end{equation}
where $s_j$ stands for  $p_j$, $M_j$, $q_j$, $N_j$.

\subsection{Nesting relations for right-hand edges}
\label{sec: 4.15}

Instead of setting up a recursion for the left-hand edges of the nested 
sub-images, we could also set up a recursion for their right-hand edges. 
If the right hand edge of the internally nested sub-image is at a renormalised 
flux value of $\tilde \phi^{\rm R}$, then equation (\ref{eq: 3.4}) 
gives 
\begin{equation}
\label{eq: 4.2.1}
\tilde \phi^{\rm R}\equiv \frac{\tilde p^{\rm R}}{\tilde q^{\rm R}}=\frac{\tilde p^{\rm L}+\tilde M}{\tilde q^{\rm L}-\tilde N}
\ .
\end{equation}
The equations describing the nesting need to be modified, because the internal or renormalised coordinate 
of a sub-image is understood to be $\phi'=0$ on the left-hand edge and $\phi'=1$ on the right-hand edge. 
Equation (\ref{eq: 4.1}) is replaced by 
\begin{equation}
\label{eq: 4.2.2}
\phi_{j+1}^{\rm R}=\frac{p_{j+1}^{\rm R}}{q_{j+1}^{\rm R}}
=\frac{p_j^{\rm L}\tilde q^{\rm R}+M_j\tilde p^{\rm R}}{q_j^{\rm L}\tilde q^{\rm R}-N_j\tilde p^{\rm R}}
=\frac{(p_j^{\rm R}-M_j)\tilde q^{\rm R}+M_j\tilde p^{\rm R}}{(q_j^{\rm R}+N_j)\tilde q^{\rm R}-N_j\tilde p^{\rm R}}
\end{equation}
where the final step uses (\ref{eq: 3.4}). Equation (\ref{eq: 4.2}) also needs to be modified. 
For a sub-image based upon a band with left- and right-hand edges $p_0^{\rm L}/q_0^{\rm R}$, 
$p_0^{\rm R}/q_0^{\rm R}$ and with Hall conductance integers $M_0$, $N_0$, we have 
\begin{equation}
\label{eq: 4.2.3}
\phi'=\frac{q_0^{\rm L}\phi-p_0^{\rm L}}{N_0\phi+M_0}
=\frac{(q_0^{\rm R}+N_0)\phi-(p_0^{\rm R}-M_0)}{N_0\phi+M_0}
\ .
\end{equation}
If we replace (\ref{eq: 2.1.1}) with this expression, the argument in section \ref{sec: 2.2} leading to
equation (\ref{eq: 2.2.8}) now gives
\begin{equation}
\label{eq: 4.2.4}
M_{j+1}=\tilde M M_j-\tilde N(p^{\rm R}_j-M_j)
\ ,\ \ \ 
N_{j+1}=\tilde M N_j+\tilde N(q^{\rm R}_j+N_j)
\ .
\end{equation}
The right-hand edges are described by an iteration in the form 
$\mbox{\boldmath$x$}^{\rm R}_{j+1}={\bf A}^{\rm R}\,\mbox{\boldmath$x$}^{\rm R}_j$
where $\mbox{\boldmath$x$}^{\rm R}_j=(p_j^{\rm R},M_j,q_j^{\rm R},N_j)$ and 
\begin{equation}
\label{eq: 4.2.5}
{\bf A}^{\rm R}=
\left(\begin{array}{cccc}
\tilde q^{\rm R}&\tilde p^{\rm R}-\tilde q^{\rm R}&0&0 \cr
-\tilde N&\tilde M + \tilde N &0&0 \cr
0&0&\tilde q^R&-(\tilde p^{\rm R}-\tilde q^{\rm R})\cr
0&0&\tilde N&\tilde M + \tilde N
\end{array}\right)
\ .
\end{equation}
Alternatively, we can write the iteration in the form of equations 
(\ref{eq: 4.4a}) and (\ref{eq: 4.4b}), with ${\bf C}^{\rm L}$ replaced by
\begin{equation}
\label{eq: 4.2.6}
{\bf C}^{\rm R}=
\left(\begin{array}{cc}
\tilde q^{\rm R} & \tilde p^{\rm R}-\tilde q^{\rm R}\cr
-\tilde N & \tilde M+\tilde N
\end{array}\right)
\ .
\end{equation}

\subsection{Scaling factors}
\label{sec: 4.2}

The eigenvalues of the matrix (\ref{eq: 4.4}) are two doubly-degenerate pairs. Noting equation
(\ref{eq: 2.1.2}) (in the form $1=\tilde q\tilde M+\tilde p\tilde N$), they are equal to
\begin{equation}
\label{eq: 4.5}
\lambda_{\pm}=\frac{(\tilde q^{\rm L}+\tilde M)}{2}\pm\sqrt{\left(\frac{\tilde q^{\rm L}+\tilde M}{2}\right)^2-1}
\ .
\end{equation}
The values of $\tilde p^{\rm L}$, $\tilde q^{\rm L}$ and $\tilde M$ describe how the nested sequences of 
sub-images are constructed. We consider a band with left edge {$\phi_0^{\rm L}=p_0^{\rm L}/q^{\rm L}_0$, 
with Hall conductance $M_0$, and its corresponding sub-image. We can recursively 
construct sub-images nested within this one, with an \lq internal' or renormalised flux at the 
left edge equal to $\phi'=\tilde p^{\rm L}/\tilde q^{\rm L}$. The Hall conductance integers of the renormalised 
Hamiltonian are $\tilde M$, $\tilde N$. In terms of the full-scale Hofstadter butterfly plot, the left edge of the 
new sub-image is $\phi^{\rm L}_1=p^{\rm L}_1/q^{\rm L}_1$. 
We can then recursively construct further sub-images in the 
same manner, with their left edges at flux values $\phi^{\rm L}_j=p^{\rm L}_j/q^{\rm L}_j$. 
The $\phi^{\rm L}_j$ may be obtained 
by iteration of (\ref{eq: 4.3}) with initial values $\mbox{\boldmath$x$}^{\rm L}_0=(p^{\rm L}_0,M_0,q^{\rm L}_0,N_0)$, 
and the corresponding right-hand edges are obtained using 
(\ref{eq: 3.4}). The bands constructed by iteration of this 
procedure converge towards a point in the original Hofstadter butterfly plot as the transformation is iterated. 
The sizes of the sub-images 
decrease with each iteration: their horizontal extent is 
\begin{equation}
\label{eq: 4.6}
\Delta \phi_j=\vert \phi^{\rm L}_j-\phi_j^{\rm R}\vert=
\bigg\vert\frac{p_j^{\rm L}q^{\rm R}_j-q^{\rm L}_jp_j^{\rm R}}{q_j^{\rm L}q_j^{\rm R}}\bigg\vert
\ .
\end{equation}
Using (\ref{eq: 3.61}), we obtain
\begin{equation}
\label{eq: 4.7}
\Delta \phi_j=\frac{1}{q_j^{\rm L}q_j^{\rm R}}
\ .
\end{equation}
As $j\to \infty$ the solutions of (\ref{eq: 4.3}) are determined 
by the eigenvalues of ${\bf A}^{\rm L}$ which have the largest magnitude. 
These will be denoted by $\lambda_\ast$. Note that, because $\tilde q^{\rm R}=\tilde q^{\rm L}-\tilde N$, 
the matrices ${\bf A}^{\rm L}$ and ${\bf A}^{\rm R}$ have the same eigenvalues. 
The scaling ratio between the sizes of successive sub-images therefore 
approaches a limit
\begin{equation}
\label{eq: 4.8}
\lim_{j\to \infty}\frac{\Delta \phi_{j+1}}{\Delta \phi_j}=\frac{1}{\lambda_\ast^2}
\end{equation}
where $\lambda_\ast$ is the larger in magnitude of the $\lambda_\pm$ 
obtained from equation (\ref{eq: 4.5}). 
The values of $M_j$ and $N_j$ also grow under iteration of the map (\ref{eq: 4.3}): 
in this case the ratio of successive values is asymptotic to the eigenvalue with the 
largest magnitude, $\lambda_\ast$: we have
\begin{equation}
\label{eq: 4.90}
\lim_{j\to \infty}\frac{M_{j+1}}{ M_j}=\lim_{j\to \infty}
\frac{N_{j+1}}{ N_j}= \lim_{j\to \infty}\frac{p^{\rm X}_{j+1}}{ p^{\rm X}_j}
=\lim_{j\to \infty}\frac{q^{\rm X}_{j+1}}{ q^{\rm X}_j}
=\lambda_\ast
\end{equation}
where, X may stand for L, c or R, so that $\phi^{\rm X}_j=p^{\rm X}_j/q^{\rm X}_j$ is the 
flux parameter at, respectively, the left hand edge, the centre, or the right-hand edge 
of the sub-image after $j$ iterations. Therefore, as we zoom 
in the (asymptotically) self-similar  sequence of sub-images,
their  flux intervals shrink by the scaling factor $\lambda_\ast^2$.  The topological 
integers $(M_j,N_j)$ and the numerator $p^{\rm X}_j$ and the denominator $q^{\rm X}_j$ of the 
rational flux $\phi^{\rm X}_j=\frac{p^{\rm X}_j}{q^{\rm X}_j}$
(with ${\rm X}\in ={\rm L},{\rm R},{\rm c}$) grow by the scaling factor $\lambda_\ast$. 
We will refer to $\lambda_\ast$ as the \emph{scaling factor} for the set of nested sub-images. 
Equations  (\ref{eq: 4.8}) and (\ref{eq: 4.90}) 
show that $\lambda_\ast$, which is a simple function of the 
integer $\tilde q^{\rm L}+\tilde M$, characterises many aspects of the self-similarities 
which are contained in the butterfly plot.

\subsection{Examples}
\label{sec: 4.3}

Figure \ref{fig: 3} shows two examples of nested sequences of sub-images. 
We consider an initial sub-image, the one shown in blue in figure \ref{fig: 1}, with 
$\phi_{\rm L}=1/3$, $k=3$, (for which $M=0$, $N=1$, 
and hence $\phi_{\rm R}=1/2$), and then build two sequences of nested sub-images of 
this.  

In the  upper panels of figure \ref{fig: 3} we follow a nested sequence of sub-images 
of the red sub-image of figure \ref{fig: 1}, with $\phi_{\rm L}=1/3$, $k=3$ 
(for which $M=0$, $N=1$, implying $\phi_{\rm R}=1/2$) , which has the same relation to 
the first sub-image as that sub-image has to the entire Hofstadter plot. We iterate equations 
(\ref{eq: 4.3}) and (\ref{eq: 4.4}) with the initial conditions 
\begin{equation}
\label{eq: 4.9}
\mbox{\boldmath$x$}^{\rm L}_0=(p^{\rm L}_0,M_0,q^{\rm L}_0,N_0)=(1,0,3,1)
\ .
\end{equation}
We then use the following values in equation (\ref{eq: 4.4}) 
to determine the matrix ${\bf A}^{\rm L}$:
\begin{equation}
\label{eq: 4.10}
(\tilde p^{\rm L},\tilde M,\tilde q^{\rm L},\tilde N)=(1,0,3,1)
\ .
\end{equation}
Iteration of equation (\ref{eq: 4.4}) then gives the following for the first three iterates:
\begin{eqnarray}
\label{eq: 4.11}
\mbox{\boldmath$x$}^{\rm L}_1=(p^{\rm L}_1,M_1,q^{\rm L}_1,N_1)&=&(3,-1,8,3)
\nonumber \\
\mbox{\boldmath$x$}^{\rm L}_2=(p^{\rm L}_2,M_2,q^{\rm L}_2,N_2)&=&(8,-3,21,8)
\nonumber \\
\mbox{\boldmath$x$}^{\rm L}_3=(p^{\rm L}_3,M_3,q^{\rm L}_3,N_3)&=&(21,-8,55,21)
\ .
\end{eqnarray}
The corresponding right-hand edges and centre points 
are then obtained using the values of $M_j$ and $N_j$ in equation (\ref{eq: 4.11}), 
together with equations (\ref{eq: 3.4}) and (\ref{eq: 3.6}).  
We find the following values for the three generations:
\begin{eqnarray}
\label{eq: 4.12}
{\rm First\ generation}&=&\phi_{\rm L}=3/8,\ \phi_{\rm R}=2/5,\ \phi_{\rm c}=5/13
\nonumber \\
{\rm Second\ generation}&=&\phi_{\rm L}=8/21,\ \phi_{\rm R}=5/13,\ \phi_{\rm c}=13/34
\nonumber \\
{\rm Third\ generation}&=&\phi_{\rm L}=21/55,\ \phi_{\rm R}= 13/34,\ \phi_{\rm c}=55/89
\ .
\end{eqnarray}
In this example,  eigenvalues are $\lambda_\pm=(3\pm\sqrt{5})/2$, so that the ratio 
of the sizes of the nested sub-images approaches 
$((3+\sqrt{5})/2)^2=(7+3\sqrt{5})/2$. 

In the lower sequence of figure \ref{fig: 3} we follow a nested sequence of sub-images 
of the blue sub-image from figure \ref{fig: 1}, this time 
based upon a nested sequence of sub-images for which the \lq internal' 
or renormalised coordinate of the left-hand edge is $\tilde \phi=3/5$, with band index $k=3$. 
For this band the Hall conductance integers are $\tilde M=-1$, $\tilde N=2$, 
implying that the right-hand edge has internal coordinate $\tilde \phi_{\rm R}=2/3$. 
Accordingly, we iterate equations (\ref{eq: 4.3}) and (\ref{eq: 4.4})
with the same initial conditions 
\begin{equation}
\label{eq: 4.9}
\mbox{\boldmath$x$}^{\rm L}_0=(p^{\rm L}_0,M_0,q^{\rm L}_0,N_0)=(1,0,3,1)
\end{equation}
and we use the following values in equation (\ref{eq: 4.4}) to determine the matrix ${\bf A}^{\rm L}$:
\begin{equation}
\label{eq: 4.10}
(\tilde p^{\rm L},\tilde M,\tilde q^{\rm L},\tilde N)=(3,-1,5,2)
\ .
\end{equation}
Iteration of equation (\ref{eq: 4.4}) then gives the following for the first three iterates:
\begin{eqnarray}
\label{eq: 4.11}
\mbox{\boldmath$x$}^{\rm L}_1=(p^{\rm L}_1,M_1,q^{\rm L}_1,N_1)&=&(5,-2,12,5)
\nonumber \\
\mbox{\boldmath$x$}^{\rm L}_2=(p^{\rm L}_2,M_2,q^{\rm L}_2,N_2)&=&(19,-8,45,19)
\nonumber \\
\mbox{\boldmath$x$}^{\rm L}_3=(p^{\rm L}_3,M_3,q^{\rm L}_3,N_3)&=&(71,-30,168,71)
\ .
\end{eqnarray}
From these vectors we extract the following values for the left-hand edges
$\phi_j^{\rm L}$ of a nested sequence of sub-images: $\phi^{\rm L}_1=5/12$, $\phi^{\rm L}_2=19/45$, 
$\phi^{\rm L}_3=71/168,\ldots $. We predict that there will 
be generations of nested sub-images with the following edges and centres
\begin{eqnarray}
\label{eq: 4.12}
{\rm First\ generation}&=&\phi_{\rm L}=1/3,\ \phi_{\rm R}=1/2,\ \phi_{\rm c}=2/5
\nonumber \\
{\rm Second\ generation}&=&\phi_{\rm L}=5/12,\ \phi_{\rm R}=3/7,\ \phi_{\rm c}=8/21
\nonumber \\
{\rm Third\ generation}&=&\phi_{\rm L}=19/45,\ \phi_{\rm R}=11/26,\ \phi_{\rm c}=30/71
\ .
\end{eqnarray}
In this case the eigenvalues are $\lambda_\pm=2\pm\sqrt{3}$, so that the ratio 
of the sizes of the nested sub-images approaches $(2+\sqrt{3})^2=7+4\sqrt{3}$.

\subsection{A simplified recursion}
\label{sec: 4.4}

The recursions (\ref{eq: 4.3}) and (\ref{eq: 4.4}) describing the nesting 
of sub-images are coupled equations in $\phi^{\rm L}_j$ and the Hall conductances integers, $M_j$ and $N_j$. 
We noticed that, in the special 
case where we the parameters of the initial sub-images and those of the subsequent nesting 
are the same (that is, when $\{\tilde p^{\rm L},\tilde q^{\rm L},\tilde M,\tilde N\}=\{p^{\rm L}_0,q^{\rm L}_0,M_0,N_0\}$), the 
flux parameters $\phi_j$ satisfy a very simple recursion, which does not require information 
about the Hall conductance integers $M_j$ and $N_j$. In the special case 
where
\begin{equation}
\label{eq: 4.13a}
\mbox{\boldmath$x$}^{\rm L}_0=(p^{\rm L}_0,M_0,q^{\rm L}_0,N_0)=(\tilde p^{\rm L},\tilde M,\tilde q^{\rm L},\tilde N)
\ .
\end{equation}
we find that the recursion of the $\phi_j^{\rm L}$ is given by
\begin{equation}
\label{eq: 4.13}
\phi_{j+1}=\frac{\tilde p+\tilde M\phi_j}{\tilde q-\tilde N\phi_j}
\ .
\end{equation}
This is simpler than (\ref{eq: 4.1}) because it uses the \emph{fixed} Hall conductances $\tilde M$ and 
$\tilde N$, rather than $M_j$ and $N_j$, which depend of the index of the iteration, $j$.  
The form of equation (\ref{eq: 4.13}) bears a marked similarity to (\ref{eq: 3.3}), but we cannot deduce it directly 
from that equation. Instead we must show how it arises from equations (\ref{eq: 4.3}) and (\ref{eq: 4.4})
in the special case where (\ref{eq: 4.13a}) is satisfied.

We can obtain $\phi^{\rm L}_j=p^{\rm L}_j/q^{\rm L}_j$ from the first and third coefficients of 
$\mbox{\boldmath$x$}^{\rm L}_j=({\bf A}^{\rm L})^j\mbox{\boldmath$x$}^{\rm L}_0$. Note that, in the special 
case that we consider, 
\begin{equation}
\label{eq: 4.14}
{\bf A}^{\rm L}\mbox{\boldmath$x$}^{\rm L}_0=(\tilde q^{\rm L}
+\tilde M)\mbox{\boldmath$x$}^{\rm L}_0-\mbox{\boldmath$y$}_0
\ ,\ \ \ 
\mbox{\boldmath$y$}_0=(0,1,1,0)
\ ,\ \ \ 
{\bf A}^{\rm L}\mbox{\boldmath$y$}_0=\mbox{\boldmath$x$}_0
\ .
\end{equation}
and hence deduce that 
\begin{equation}
\label{eq: 4.15}
\mbox{\boldmath$x$}^{\rm L}_j\equiv ({\bf A}^{\rm L})^j\mbox{\boldmath$x$}^{\rm L}_0=
\alpha_j\mbox{\boldmath$x$}^{\rm L}_0+\beta_j\mbox{\boldmath$y$}_0
\ .
\end{equation}
Using (\ref{eq: 4.14}) we find that the coefficients $\alpha_j$, $\beta_j$ satisfy
\begin{equation}
\label{eq: 4.16}
\left(\begin{array}{c}
\alpha_{j+1} \cr
\beta_{j+1}
\end{array}\right)
={\bf a}
\left(\begin{array}{c}
\alpha_j \cr
\beta_j
\end{array}\right)
\ ,\ \ \ 
{\bf a}=\left(\begin{array}{cc}
\tilde q+\tilde M & 1 \cr
-1 & 0
\end{array}\right)
\end{equation}
From (\ref{eq: 4.14}) and (\ref{eq: 4.15}) we deduce that $p_j=\tilde p\alpha_j$ and $q_j=\tilde q\alpha_j+\beta_j$, 
so that (\ref{eq: 4.16}) can be expressed as a recursion of $\{p_j^{\rm L},q_j^{\rm L}\}$, in the form
\begin{equation}
\label{eq: 4.17}
\left(\begin{array}{c}
p^{\rm L}_{j+1} \cr
q^{\rm L}_{j+1}
\end{array}\right)
=\left(\begin{array}{cc}
\tilde M & \tilde p \cr
-\tilde N & \tilde q
\end{array}\right)
\left(\begin{array}{c}
p^{\rm L}_j \cr
q^{\rm L}_j
\end{array}\right)
\ .
\end{equation}
Equation (\ref{eq: 4.13}) then follows immediately.
 It is easy to check that equation (\ref{eq: 4.13}) reproduces
the left-hand band edges in the upper panel of figure \ref{fig: 3}. It does not reproduce the values of 
$\phi^{\rm L}_j$ for the lower panel, because these do not satisfy (\ref{eq: 4.13a}).

In the special case where (\ref{eq: 4.13a}) applies, we can determine the accumulation point $\phi_\ast$ 
of the nesting process as the fixed point of equation (\ref{eq: 4.13}). Setting $\phi_{j+1}=\phi_j=\phi_\ast$, we 
find
\begin{equation}
\label{eq: 4.18}
\phi_\ast=\frac{1}{2\tilde N}\left[(\tilde q-\tilde M)\pm \sqrt{(\tilde q+\tilde M)^2-4}\right]
\end{equation}
(with the sign chosen so that $0<\phi_\ast<1$).
We remark that the sequences $\phi^{\rm L}_j$, $\phi^{\rm R}_j$ bracket this accumulation 
point, and that $\phi^{\rm c}_j$ is another sequence that converges towards it. In the general case 
none of these sequences corresponds to the continued fraction representation of $\phi_\ast$.

\section{Chains of sub-images}
\label{sec: 5}

The arguments of section \ref{sec: 3} show that each of the $q_0$ 
bands of the spectrum at $\phi=p_0/q_0$ is potentially an edge of 
two different sub-images: setting $\phi'=\pm 1$, we see that the other 
edges of these two connected sub-images are at 
\begin{equation}
\label{eq: 5.1}
\phi_\pm=\frac{p_0\pm M_0}{q_0\mp N_0}
\ .
\end{equation}
These two values can themselves be the edges of further sub-images.
Because $N_0$ and $M_0$ are constant so long as gaps do not close, 
by iteration we have a sequence of edges of a connected chain of sub-butterflies, for which 
the values of $\phi$ are
\begin{equation}
\label{eq: 5.2}
\phi_j=\frac{p_0+j M_0}{q_0-j N_0}
\end{equation}
where $j$ is a positive or negative integer. 

The sub-images are constructed by taking a band at $\phi_0=p_0/q_0$ and 
increasing $\phi$ until the renormalised value $\phi'$ is equal to unity, at which point the spectrum 
of the renormalised Hamiltonian is, once again, a single band. 
The results in \cite{Wil87} do not guarantee that extrapolation to $\phi'=1$ is possible, 
and one way in which the procedure could fail is if the gaps which exist in the spectrum when
$\phi'=0$ close up as $|\phi'|$ increases. We find, for a given value of $\phi_0=p_0/q_0$, 
that it is always possible to construct a sub-image of the butterfly plot to either the right 
(setting $\phi'=+1$) or the left, ($\phi'=-1$) without gaps in the spectrum closing, 
but not always both.  We refer to sub-images which have \emph{both} upper and lower 
gaps open at \emph{both} sides as \emph{open} sub-images. 
Other sub-images will be termed \emph{closed}. 

In cases where the gaps close, we find empirically that this occurs 
when $|\phi'|=1$. As an example, consider the case of the centre band for $\phi_0=1/3$ 
(for this band $M=1$, $N=-2$). Applying (\ref{eq: 5.1}) we find $\phi_+=2/5$ and $\phi_-=0$. 
There is an open sub-image for the centre band which is bounded by $\phi_{\rm L}=1/3$ 
and $\phi_{\rm R}=2/5$, but if we consider how the centre band for $\phi_{\rm R}=1/3$ 
evolves as we approach $\phi_{\rm L}=0$, we see that the gaps at both the top and the bottom 
of the band close up as we approach $\phi=0$. 
There are other examples where only one of the gaps closes: for example let us consider the 
uppermost band when $\phi_0=2/5$ (which has $M=1$, $N=-2$). In this case $\phi_+=3/7$ 
and $\phi_-=1/3$. There is an open sub-image containing the uppermost band 
bounded by $\phi_{\rm L}=2/5$ and $\phi_{\rm R}=3/7$, but if we extend the uppermost 
band from $\phi_{\rm R}=2/5$ towards $\phi_{\rm L}=1/3$, we find the that the lower 
gap closes, so that we have a closed sub-image.

In cases where the sub-image is \lq open', the formulae for predicting the $\phi $ 
value of one edge from the other can be used reciprocally: we can determine $\phi_{\rm R}$
from $\phi_{\rm L}$ by taking the positive sign in (\ref{eq: 5.1}), or $\phi_{\rm L}$ 
from $\phi_{\rm R}$ by taking the negative sign. If, however, the sub-image has a closed 
gap at the right-hand side, while it is possible to compute $\phi_{\rm R}$ from $\phi_{\rm L}$, 
if we start from $\phi_{\rm R}$, the values of the Hall conductance integers are 
changed by the additional component of the spectrum which does not form part of the 
band at $\phi_{\rm L}$.

It is also possible for a \lq closed' edge to be shared between multiple sub-images.
An example of this is the set of centre bands at a sequence of values $\phi_{\rm R}=1/(2n+1)$,
with $n=1,2,\ldots$. Here $M=1$ and $N=-2n$. If we apply equation (\ref{eq: 5.1}) with the 
negative sign, we find that the left hand edge is $\phi_{\rm L}=0$ in every case. 
We find that the left-hand edge of each sub-image based upon the 
centre band at $\phi_{\rm R}=1/(2n+1)$ is the entire spectrum at $\phi=0$, 
for all integer $n>0$. The reason why this band can be the left hand edge of an infinite number
of different sub-images is that, as $\phi\to 0$, an infinite number of bands accumulate 
at both the upper and lower edges of the spectrum.

For every band at a rational value of $\phi$, we can attempt to construct a connected 
chain of sub-images extending in either direction using equation (\ref{eq: 5.2}). 
We find that the chain terminates in one direction, due to encountering an edge where 
the gaps close. The chain extends to infinite values of $|j|$ in the other direction. 
Note that $\phi_j$, defined by (\ref{eq: 5.2}), approaches $-M_0/N_0$ as $|j|\to \infty$,
so that the chain of sub-images ends at an accumulation point. For example, every 
centre band for $\phi=1/(2n+1)$ (with $n=1,2,\ldots$, having Hall integers $M_0=1$ 
and $N_0=-2n$) has a chain of sub-images extending 
to the right, with edges $\phi_j=(j+1)/[2n(j+1)+1]$, end at an accumulation point at 
$\phi=1/2n$. 

Table \ref{T1} lists parameters of four different examples of chains, 
which are illustrated in figure 4. The chains are denoted 
by a label $C_{\frac{p_{\rm cl}}{q_{\rm cl}}\rightarrow \frac{p_{\rm ac}}{q_{\rm ac}},k\pm}$ 
in which $p_{\rm cl}/q_{\rm cl}$ is the closed edge and $p_{\rm ac}/q_{\rm ac}$ is the accumulation 
point, and $k$ labels the band to which the accumulation point attaches, with $\pm$
indicating whether the attachment is to the top $+$, or bottom, $-$. The edges of the sub-images 
forming the chain are denoted by $\phi_j$, with $j=0$ being the edge for which a gap 
closes, and $j=1,2,\ldots$ being labels of the open edges.

\begin{table}
\begin{tabular}{| c | c |  c | c |}
\hline
$C_{\frac{p_{\rm cl}}{q_{\rm cl}}   \rightarrow \frac{p_{\rm ac}}{q_{\rm ac}},k\pm}$& 
$\frac{p_j}{q_j}$ \,\, &   $(M_0,N_0)$\,\, & {\rm colour}  \\ 
\hline
$C_{\frac{1}{2}   \rightarrow 0,1+}$ \,\,  & $\frac{1}{j+1}$ \,\, & $(0,1) $\,\, & {\rm red}  \\ 
\hline
$C_{0   \rightarrow   \frac{1}{2},1+,2-}$  \,\, & $\frac{j-1}{2j-1}$\,\, &  $(1, -2) $\,\, & {\rm blue}  \\ 
\hline
$C_{\frac{1}{3}\rightarrow  \frac{1}{4},2+,3-}$  \,\, & $ \frac{1+j}{3+4j}$\,\, &  $(-1,4 )$\,\, & {\rm   green}\\ 
\hline
$C_{\frac{1}{4} \rightarrow  \frac{1}{3},2+}$\,\, & $\frac{j+1}{3j+4} $\,\, &  $(1,-3)$\,\, & {\rm purple} \\ 
\hline
\hline
\end{tabular}
\caption{ 
Four infinite chains of sub-images 
(labelled by $C_{\frac{p_{\rm cl}}{q_{\rm cl}}\rightarrow
\frac {p_{\rm ac}}{q_{\rm ac}},k\pm}$, where $\frac{p_{\rm cl}}{q_{\rm cl}}$
is the open edge, and $\frac {p_{\rm ac}}{q_{\rm ac}}$ is the accumulation point, attaching 
to band $k$, at the the upper, $+$, or lower edge, $-$, edge). 
We list the flux parameter of one edge of a sub-image, $p_j/q_j$, with $j=0$ giving the closed edge 
and $j=1,2,\ldots$ giving the open edges, the values of the Hall 
integers for the chain, $(M_0,N_0)$, and the colour used to highlight the chain in figure 4.}
\label{T1}
\end{table}

\begin{figure}
\label{fig: 4}
\includegraphics[width=0.75\textwidth]{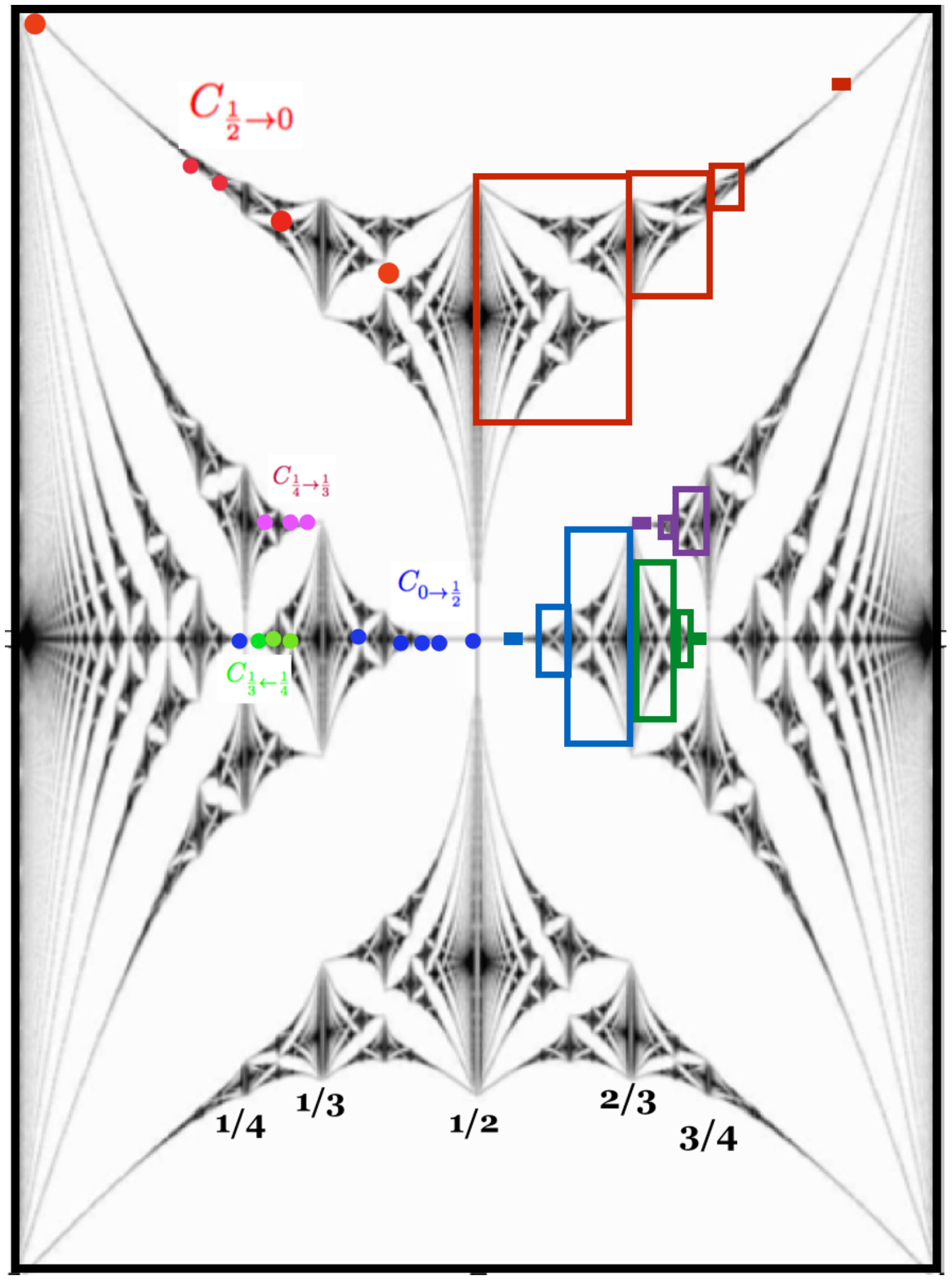}
\caption{
(Colour online). Illustrating examples of four chains (colour coded in red, blue, green and purple) of sub-images, 
specified in table \ref{T1}.  Exploiting left-right symmetry of the graph, chains are shown 
in two ways. On the left, each member of the chain is shown with a dot at the centre 
of the sub-image and on the right, the  sub-images are shown in the boxes.}
\end{figure}

\section{Summary}
\label{sec: 6}

A striking feature of Hofstadter's butterfly is the fact that its interior can be dissected into small, 
distorted images of the entire plot. These sub-images are a microcosm of the butterfly plot.

In this work we have described how every band of the spectrum for rational $\phi$ can be taken to be 
an edge of at least one sub-image, and we have shown how the other edge or edges can be 
determined. The centre and the edges are shown to be neighbouring fractions in the Farey tree and 
the equations relating them depend upon the quantised Hall conductance integers, $M$ and $N$.

We have also analysed two ways in which these objects can be interrelated, namely 
by being recursively nested, or by forming chains. Each sub-image is described by four integers:
$p$ and $q$ specify the flux ratio (on one of the edges), and $M$, $N$ specify its associated quantised 
Hall conductance. Both the nesting and concatenation relationships between sub-images are represented 
by simple algebraic operations on the set $\{p,M,q,N\}$. Equations (\ref{eq: 4.4a}) and (\ref{eq: 4.4b}) 
show that nesting relationship may be represented by multiplication of $2\times 2$ unimodular matrices 
composed from these numbers. Equation (\ref{eq: 5.2}) shows that the concatenation relation corresponds 
to a simple additive relation.

The nesting relationship leads to 
a the derivation of exact expressions for the scaling factors describing self-similarity.
The sub-images are described by rational fluxes at every step of the recursion, but 
their asymptotic scaling factors and accumulation points (equations (\ref{eq: 4.5}) and (\ref{eq: 4.18}) respectively)
are irrational numbers, obeying simple integer-coefficient quadratic equations. It is surprising 
that the scaling factors do not require the solution of matrix eigenvalue equations. 
It can also be noted that the set of quadratic numbers which can arise from equation 
(\ref{eq: 4.5}) does not include the golden mean, $(1+\sqrt{5})/2$, which features so 
extensively in the literature on Harper's equation, reviewed in \cite{Sat16}. 
The chains are infinite in one direction, with successively smaller sub-butterflies 
reaching an accumulation point. In contrast to the accumulation points of the nesting process, 
the chains end at a rational value of $\phi$. 

{\bf Acknowledgements}. IIS thanks the Department of Mathematics and Statistics at the Open University, where 
this work was started. MW thanks the Physics Department of George Mason University 
its hospitality.

\end{document}